\documentclass{PoS}

\usepackage{subfigure}
\usepackage{psfrag}

\newlength{\halffigwidth}
\title{The lattice infrared Landau gauge gluon propagator: the infinite volume limit}

\ShortTitle{The lattice infrared Landau gauge gluon propagator: the infinite volume limit}

\author{O. Oliveira$^1$, \speaker{P. J. Silva} $^{1,2}$\\
\llap{$^1$} Centro de F\'{i}sica Computacional, Universidade de Coimbra\\
\llap{$^2$} School of Physics and Astronomy, The University of Edinburgh\\
        E-mail: \email{orlando@teor.fis.uc.pt}, \email{psilva@ph.ed.ac.uk} }

\abstract{We report on the infrared behaviour of the lattice
gluon propagator. The bounds on D(0) and their behaviour with the
lattice volume are investigated, together with the full propagator.
Moreover, a study of the gluon propagator
using different lattice spacings is carried out.}

\FullConference{The XXVII International Symposium on Lattice Field Theory\\
                 July 26-31, 2009\\
                 Peking University, Beijing, China}

\begin{document}

\section{Introduction and motivation}

Despite the huge efforts of the last years, up to now, we still do not have a clear understanding about the reasons why quarks and gluons remain confined in colour singlets.

The infrared properties of the gluon and ghost propagators in Landau  
gauge have been related with possible gluon confinement mechanisms. In  
this sense, the computation of the propagators can help in the  
understanding of the QCD confinement mechanism [1].

\subsection{Gluon confinement in Landau gauge}

The so-called Kugo-Ojima scenario (KO) \cite{KuOj} and the Gribov-Zwanziger horizon condition (GZ) \cite{Gribov,Zw91} impose restrictions on the infrared behaviour of the Landau gauge gluon and ghost propagators,
\begin{eqnarray}
  D^{ab}_{\mu\nu} (q) & = & \delta^{ab} \, 
           \left( \delta_{\mu\nu} \, - \, \frac{q_\mu q_\nu}{q^2} \right)
           \, D(q^2) \, ,  \\
  G^{ab} (q) & = & - \delta^{ab} G(q^2) \, ,
\end{eqnarray}

The GZ mechanism requires $D(0) = 0$ (which implies maximal violation of reflection positivity) and an enhanced ghost propagator, relative to the perturbative function. The KO confinement mechanism demands $1 / (q^2 G(q^2)) =  0$ in the limit $q \rightarrow 0$. From the point of view of the KO and GZ confinement mechanisms, the requirements on $D(0)$ and $G(0)$ are necessary conditions and its violation immediatly rules out these scenarios.

The infrared behaviour of these propagators can not be studied using perturbative methods. One should rely on non-perturbative methods. One possibility is the use of Dyson-Schwinger equations, which are an infinite tower of coupled nonlinear equations for the QCD Green's functions. Although they can provide an analytical solution in the infrared, the computation of the gluon and ghost propagators within DSE requires defining a truncation scheme and parameterizing some of the QCD vertices.

For the Landau gauge, recent solutions can be classified in two
 categories: solutions with a finite $D(0)$, which are not compatible with the KO or GZ confining mechanism (see \cite{Ag08} and references therein)
and solutions which do not rule out the two confining mechanisms (see \cite{Review} for a review).  For the later class of solutions, they predict a pure power law behaviour in the infrared, 
\begin{eqnarray}
  Z( q^2 ) ~ = ~ q^2 \, D( q^2) & = &
         \omega \left( {q^2}\right)^{2 \kappa} ,
  \label{Zdse} \\
  F( q^2 ) ~ = ~ q^2 \, G( q^2) & = &
           \omega^\prime \left( {q^2} \right)^{- \kappa} ;
  \label{Fdse}
\end{eqnarray}
 with $\kappa = 0.595$, which, for the zero momentum, implies a null (infinite) gluon (ghost) propagator.

Lattice QCD simulations provide another method of study these quantities non perturbatively, although one should care about finite volume and finite lattice spacing effects. Despite efforts in large simulations in recents years (see, for example, \cite{berlin09,Cuc0712}), lattice results show a finite non-vanishing zero momentum gluon propagator, with no sign for a turnover in the infrared region. 

In this paper we present an update on our study of Cucchieri-Mendes bounds \cite{Cuc0712} in SU(3) lattice gauge theory \cite{bounds_su3}.

\subsection{Cucchieri-Mendes bounds}

The Cucchieri-Mendes bounds \cite{Cuc0712} provide upper and lower bounds for the zero momentum gluon propagator of lattice Yang-Mills theories in terms of the average value of the gluon field. In particular, they relate the gluon propagator at zero momentum $D(0)$ with
\begin{equation}
   M(0) ~ = ~ \frac{1}{d \left( N^2_c - 1 \right)} \sum_{\mu, a} \left| A^a_\mu (0) \right| \, ,
\end{equation}   
where $d$ is the number of space-time dimensions, and $N_c$ the number of colors. In the above equation, $A^a_\mu (0)$ is the $a$ color component of the gluon field at zero momentum, defined by
\begin{equation}
  A^a_\mu (0) ~ = ~ \frac{1}{V} \sum_x A^a_\mu (x)
\end{equation}  
where $A^a_\mu (x)$ is the $a$ color component of the gluon field in the real space. According to
\cite{Cuc0712}, the $D(0)$ is related with $M(0)$ by
\begin{equation}
  \langle M(0) \rangle^2 ~ \le ~ \frac{D(0)}{V} ~ \le d \left(N^2_c - 1\right) \langle M(0)^2 \rangle \, .
  \label{bounds}
\end{equation}  
In the last equation $\langle ~ \rangle$ means Monte Carlo average over gauge configurations. The
bounds in equation (\ref{bounds}) are a direct result of the Monte Carlo approach. 
The interest on these bounds comes from allowing a scaling analysis which can help understanding
the finite volume behaviour of $D(0)$: assuming  that each of the terms in inequality (\ref{bounds}) scales with the volume according to $ A / V^\alpha$, the simplest possibility 
and the one considered in \cite{Cuc0712}, an $\alpha > 1$  for $ \langle M(0)^2 \rangle$ clearly indicates that $D(0) \rightarrow 0$ as the infinite volume is approached. In this sense, this scaling analysis allows to investigate the behaviour of $D(0)$ in the infinite volume limit. 

 For the SU(2) Yang-Mills theory \cite{Cuc0712},  the results
show a $D(0)=0$ for the two dimensional theory, but a $D(0) \ne 0$ for three and four dimensional formulations.
Our analysis for SU(3) favors a $D(0) = 0$ result.

\section{Cucchieri-Mendes bounds in SU(3) gauge theory}

We have studied the Cucchieri-Mendes bounds for SU(3) gauge theory with two values of the gauge coupling: $\beta=6.0$ \cite{bounds_su3} and $\beta=5.7$.

\subsection{Scaling analysis for $\beta=6.0$}

For $\beta=6.0$, we used the lattice setup described in table \ref{beta60setup}. The differences with \cite{bounds_su3} being that we now use a new ensemble for $24^4$, and we also use preliminary data for $80^4$ lattices.

\begin{table}[b]
\begin{center}
\begin{tabular}{ccccccccc}
\hline
$L^4$   &  $16^4$ & $20^4$ & $24^4$ & $28^4$ & $32^4$  & $48^4$ & $64^4$ & $80^4$ \\
L(fm)  &   1.63  & 2.03   & 2.44   & 2.84  & 3.25    &  4.88  & 6.50 & 8.13 \\
$\# conf$ &   52    & 72     &  60$^{*}$& 56    & 126     &  104   & 120 & 18$^{*}$ \\
\hline
\end{tabular}
\caption{Lattice setup for $\beta=6.0$ configurations. The lattice spacing is $a=0.1016(25)$fm. * stands for new data, comparing with \cite{bounds_su3}.\label{beta60setup}}
\end{center}
\end{table}

\begin{figure}[t]
\vspace*{0.2cm}
\begin{center}
\includegraphics[width=0.5\textwidth]{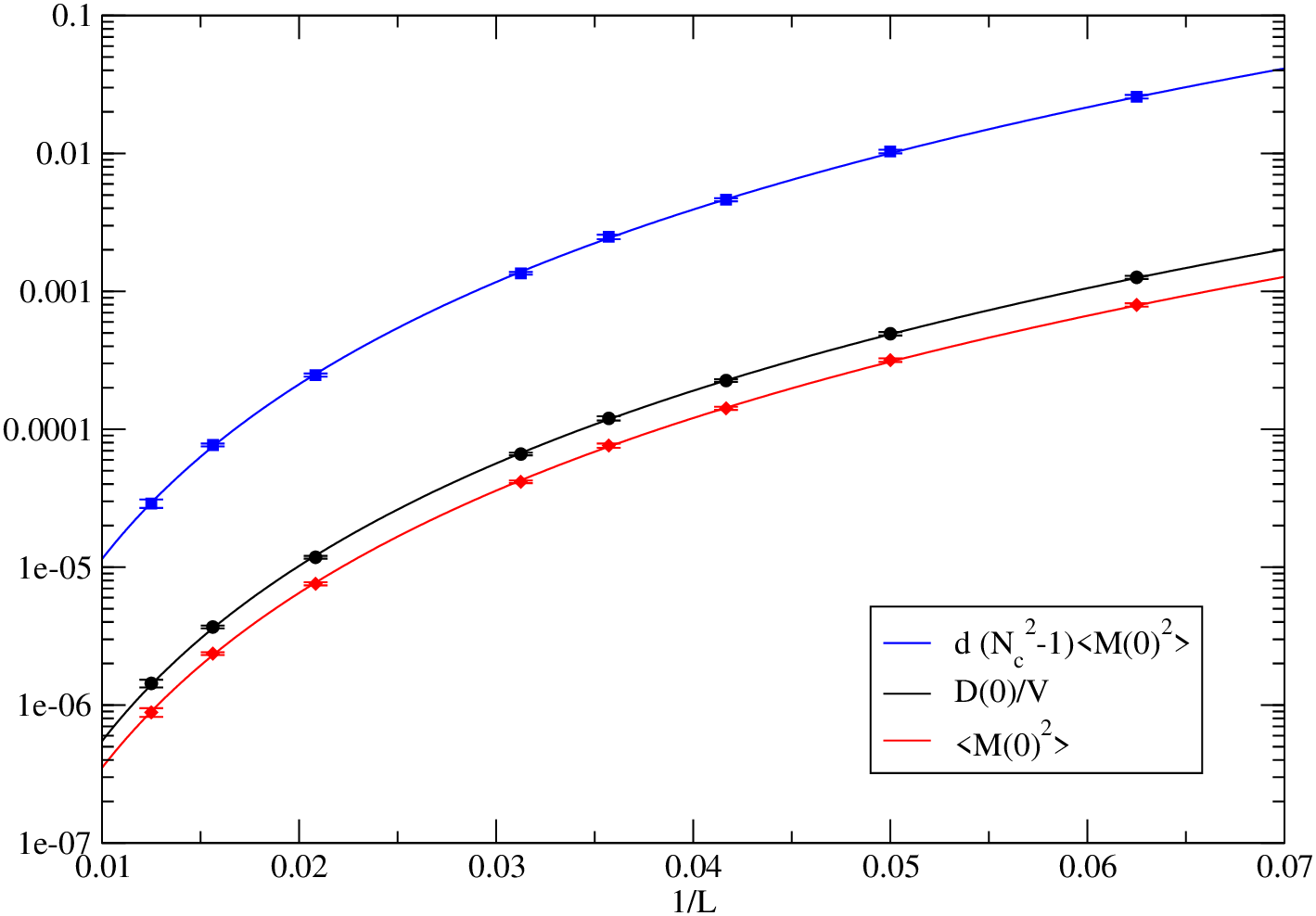}
\end{center}
\caption{Cucchieri-Mendes bounds for the lattice data at $\beta=6.0$.}
\label{boundsb60}
\end{figure}

In figure \ref{boundsb60}, we show $\langle M(0) \rangle^2$, $D(0)/V$ and 
$d (N^2_c - 1) \langle M(0)^2 \rangle$, together with the fits to $A / V^\alpha$. 

\begin{table}[b]
\begin{center}
\begin{tabular}{rrrr}
\hline
                           &  $\omega$  & $\alpha$    & $\chi^2_{\nu}$ \\
\hline
  $\langle M(0) \rangle$   &  9.66(37)  & 0.5265(27)  & 0.69  \\
  $D(0)/V$                 & $151\pm11$    & 1.0551(51)  & 0.55 \\
 $N_d(N_c^2-1) \langle M(0)^2 \rangle$ & $2999\pm228$  & 1.0522(54)  & 0.71  \\
 \hline
\end{tabular}
\end{center}
\caption{Fits to $A/V^{\alpha}$ using lattice data at $\beta=6.0$.\label{fitspplb60}}
\end{table}
\begin{table}[b]
\begin{center}
\begin{tabular}{rrrrr}
\hline
                                       &  $\omega$   &  $\alpha$    &  $C$     & $\chi^2_{\nu}$ \\
\hline
  $\langle M(0) \rangle^2$               & $125\pm74$ & $1.148(8)$  & $28.2\pm8.9$   &  0.57 \\
  $D(0)/V$                             & $247\pm201$  & $1.17(10)$    & $46\pm13$  & 0.40  \\
  $N_d(N_c^2-1)\langle M(0)^2 \rangle$  & $5215\pm258$ & $1.18(11)$  & $1015\pm258$   & 0.60  \\
 \hline
\end{tabular}
\end{center}
\caption{Fits to $C/V+\omega V^{-\alpha}$ --- lattice data at $\beta=6.0$.\label{fitsconstb60} }
\end{table}
In table \ref{fitspplb60} we see the results of fitting the lattice data to $A/V^{\alpha}$. 
The values for $\alpha$ strongly support a vanishing zero momentum gluon propagator. However,
 the fits reported in table \ref{fitsconstb60} to $C/V+\omega V^{-\alpha}$, support a $D(0)=C\ne0$. 
So, no conclusive answer can be made from the analysis of the two fits.

 The fact that the $\alpha$'s in table \ref{fitspplb60}  are in disagreement with the 
results for SU(2) \cite{Cuc0712} can have several explanations. First of all, there can 
be differences between calculations done using different gauge groups, although recent 
studies seem to support the equivalence of the two theories \footnote{Note that a recent direct comparison between SU(2) and SU(3) gluon propagators showed a measurable difference in the infrared region --- \cite{trento}.} \cite{su23ptbr, su23aust}. 

Other possibilities for such a difference are the use of different lattice volumes and/or lattice spacing. Indeed, the physical volumes considered in the SU(2) study, up to $(27fm)^4$, 
are larger than the ones used in our SU(3) study --- up to $(8\,\rm{fm})^4$. On the other hand, the lattice 
spacing used in the SU(2) study is more than twice our lattice spacing. 

\subsection{Scaling analysis for $\beta=5.7$}

To avoid considering larger volumes, which is computationally demanding, we carried simulations with a lower $\beta$ value, in the case $\beta=5.7$ ($a=0.1838(11)$fm), keeping volumes similar to the ones considered previously --- see table \ref{beta60setup}. The new lattice setup is reported in table \ref{beta57setup}.

\begin{table}[t]
\begin{center}
\begin{tabular}{cccccccc}
\hline
$L^4$     &  $8^4$  & $10^4$ & $14^4$ & $18^4$ & $26^4$  & $36^4$ & $44^4$ \\
L(fm)     &   1.47  & 1.84   & 2.57   & 3.31  & 4.78    &  6.62  & 8.09  \\
$\# conf$ &   56    & 149    &  149   & 149   & 132     &  100   &   29   \\
\hline
\end{tabular}
\end{center}
\caption{Lattice setup for $\beta=5.7$ configurations. The lattice spacing is $a=0.1838(11)$fm.\label{beta57setup}}
\end{table}

\begin{figure}[t]
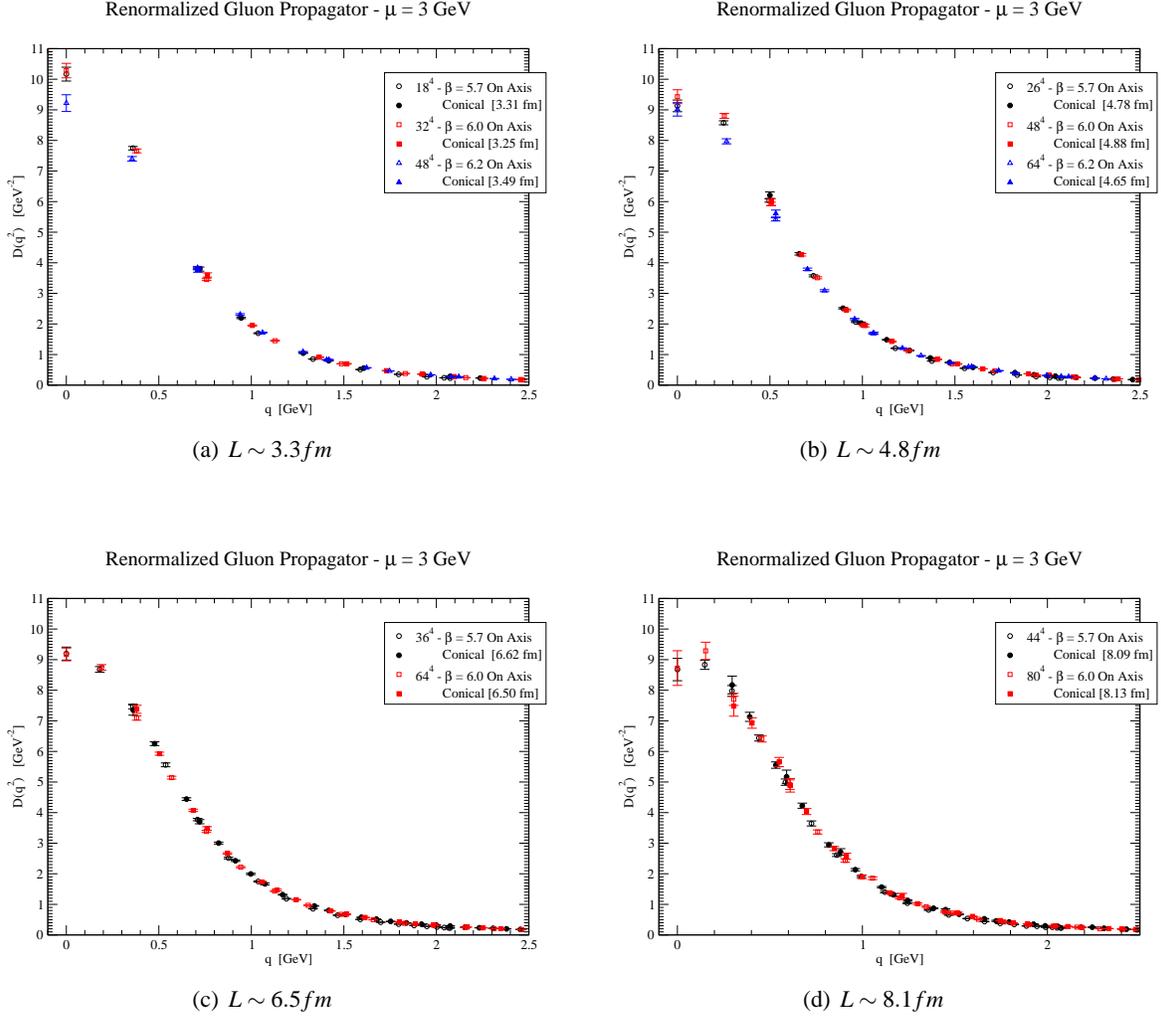

  \subfigure[$L\sim3.3fm$]{
  \begin{minipage}[b]{0.45\textwidth}
    \centering
    \includegraphics[origin=c, angle=0,width=\halffigwidth]{glue.R3GeV.V3.3fm.eps}
  \end{minipage} } \hfill
  \subfigure[ $L\sim4.8fm$ ]{
  \begin{minipage}[b]{0.45\textwidth}
    \centering
    \includegraphics[origin=c, angle=0,width=\halffigwidth]{glue.R3GeV.V4.8fm.eps}
  \end{minipage} }\vspace*{0.8cm}
  \subfigure[ $L\sim6.5fm$]{
  \begin{minipage}[b]{0.45\textwidth}
    \centering
    \includegraphics[origin=c, angle=0,width=\halffigwidth]{glue.R3GeV.V6.5fm.eps}
  \end{minipage} } \hfill
  \subfigure[ $L\sim8.1fm$]{
  \begin{minipage}[b]{0.45\textwidth}
    \centering
    \includegraphics[origin=c, angle=0,width=\halffigwidth]{glue.R3GeV.V8.1fm.eps}
  \end{minipage} }\vspace*{0.8cm}
  \caption{Comparing the gluon propagator computed with different lattice spacings at the 
same physical volume.\label{compbeta6057}}
\end{figure}

In figure \ref{compbeta6057} we can see the comparison between the propagators computed using different lattice spacings.

Note that, for the two lowest volumes, we have also plotted some data for $\beta=6.2$ 
configurations. 
Although the $\beta=5.7$ and $6.0$ data seem to be similar, there are some 
differences, in the infrared, with the propagators computed at $\beta=6.2$. 
This deserves further investigations to clarify any possible effects due to finite lattice spacing.

\begin{figure}[t]
\vspace*{0.2cm}
\begin{center}
\includegraphics[width=0.5\textwidth]{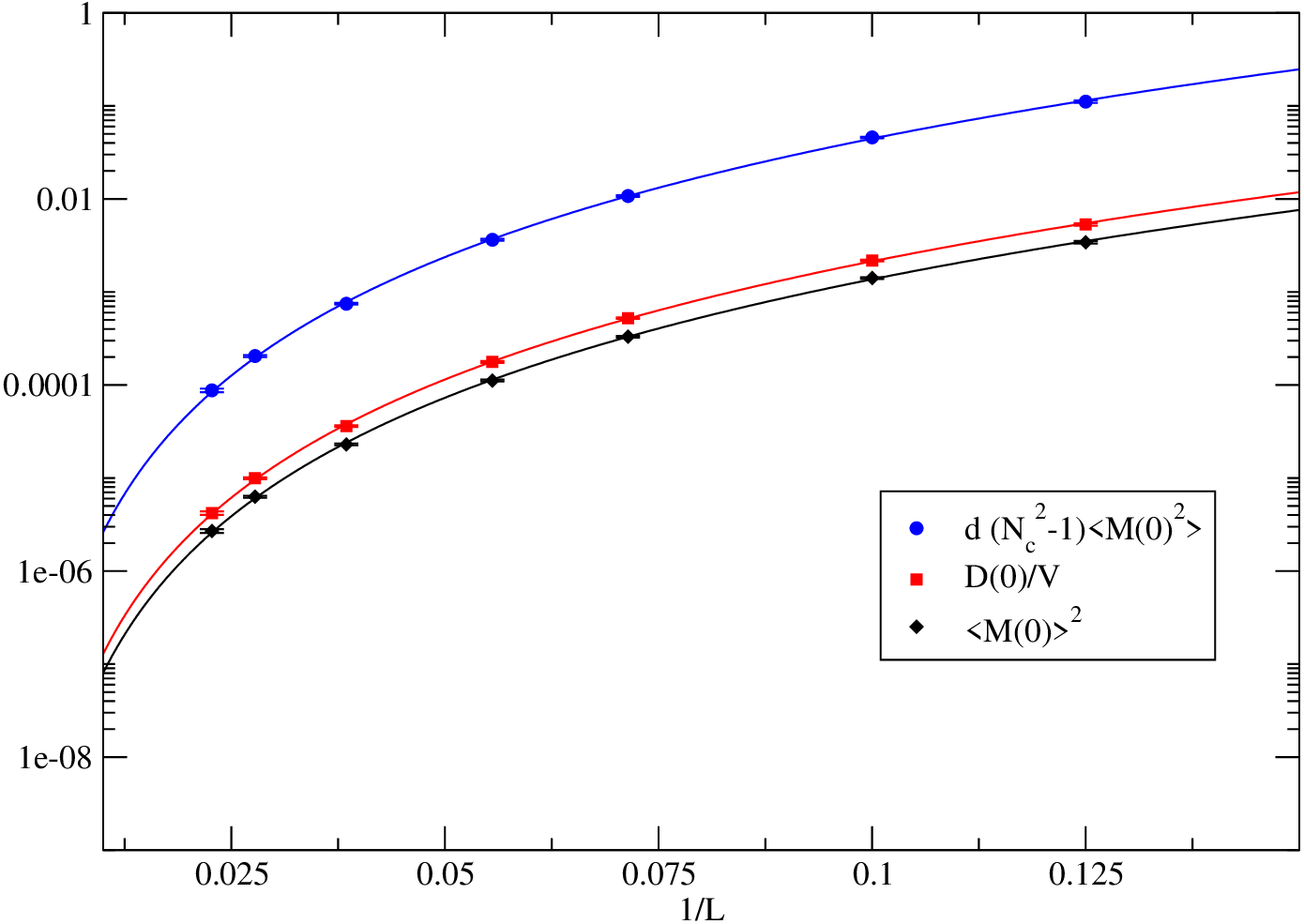}
\end{center}
\caption{Cucchieri-Mendes bounds for the lattice data at $\beta=5.7$.\label{boundsb57}}
\end{figure}

In what concerns the Cucchieri-Mendes bounds, we can see our results in figure \ref{boundsb57}. Moreover, in table \ref{fitspplb57}, we see the results of fitting\footnote{In the fits to $\beta=5.7$ data, to keep $\chi^2_{\nu}<2$ the $26^4$ lattice data had to be excluded.} the lattice data to $A/V^{\alpha}$. 

\begin{table}[b]
\begin{center}
\begin{tabular}{rrrr}
\hline
                           &  $\omega$  & $\alpha$    & $\chi^2_{\nu}$ \\
\hline
  $\langle M(0) \rangle$           &  4.73(13)   & 0.5267(25)    & 1.22 \\
  $D(0)/V$                         & $34.0\pm1.7$  & 1.0504(45)   & 0.80 \\
 $N_d(N_c^2-1) \langle M(0)^2 \rangle$ & $726\pm41$  & 1.0530(50)  & 1.08 \\
 \hline
\end{tabular}
\end{center}
\caption{Fits to $A/V^{\alpha}$ using lattice data at $\beta=5.7$.}\label{fitspplb57}
\end{table}

As in the $\beta=6.0$ case, the $\alpha$'s values strongly support a D(0)=0. However, the lattice data is also compatible with the functional form $C/V+\omega V^{-\alpha}$ --- see table \ref{fitsconstb57}. The figures presented in this table do not allow a conclusive answer, as the fit to D(0)/V allows for $D(0)=0$; however, the $C$ figure for $\langle M(0) \rangle^2$ implies a $D(0)\neq0$.

\begin{table}[b]
\begin{center}
\begin{tabular}{rrrrr}
\hline
                                       &  $\omega$   &  $\alpha$    &  $C$     & $\chi^2_{\nu}$ \\
\hline
  $\langle M(0) \rangle^2$               & $22\pm8$ & $1.13(10)$  & $7.0\pm3.7$   &  1.40 \\
  $D(0)/V$                             & $32\pm13$  & $1.063(91)$    & $3\pm22$  & 1.07   \\
  $N_d(N_c^2-1)\langle M(0)^2 \rangle$  & $688\pm189$ & $1.117(90)$  & $212\pm135$   & 1.29  \\
 \hline
\end{tabular}
\end{center}
\caption{Fits to $C/V+\omega V^{-\alpha}$ --- lattice data at $\beta=5.7$. \label{fitsconstb57}}
\end{table}

\section{Conclusions and future work}

In this article we presented a study of Cucchieri-Mendes bounds in SU(3) pure gauge theory, using two different values of lattice spacing. Fitting the data to a pure power law in the volume $A/V^{\alpha}$, one finds always $D(0)=0$. Note that, although this result is in disagreement with the SU(2) study \cite{Cuc0712}, it is in agreement with our study presented in \cite{ratios}, where we computed a $\kappa\sim0.53$. This value for $\kappa$ implies a vanishing zero momentum gluon propagator. 

However, the use of more general ansatze for the dependence with the lattice volume do not allow to take definitive conclusions. This puzzle deserves a more accurate study, together with a better understanding of lattice effects in the propagators.

\section*{Acknowledgments}

This work was supported by projects CERN/FP/83582/2008 and CERN/FP/83664/2008. Simulations have been run in supercomputer Milipeia at Universidade de Coimbra. P. J. Silva acknowlegdes support from FCT via grant SFRH/BPD/40998/2007.

\end{document}